\documentclass[intlimits,twoside,a4paper]{article}

\usepackage[cp1251]{inputenc}
\usepackage[eqsecnum]{cmpj3}

\usepackage{graphicx}
\usepackage{amsmath}
\usepackage{natbib}
\usepackage{bm}
\usepackage{epsfig}
\usepackage{xcolor}

\issue{2025}{28}{1}{13301}
\doinumber{10.5488/CMP.28.13301}

\title[Scaling properties of diblock copolymers]%
{ Scaling properties of diblock copolymers: dynamic simulations study}%

\author[K. Haydukivska]{K. Haydukivska\orcid{0000-0002-3118-7010}\thanks{Email: \email{wja4eslawa@icmp.lviv.ua}.},
}
\addresses{
\addr{label1} Institute for Condensed Matter Physics of the National Academy of Sciences of Ukraine, 1 Svientsitskii Str., 79011 Lviv, Ukraine
\addr{label2} Institute of Physics, University of Silesia, 75 Pu\l{}ku Piechoty 1, 41-500 Chorz{\'o}w, Poland
}

\date{Received January 20, 2025, in final form February 18, 2025}

\begin{document}
	
\maketitle
\begin{abstract}
The influence of monomer-monomer interactions on the scaling exponents and shape characteristics of a single polymer chain in a selective solvent is investigated using Langevin dynamics simulations. By systematically increasing the temperature of the solution, the effects of interactions between blocks on the conformational properties of the chain are explored. The results demonstrate that longer-range interactions cause a transition of a polymer similar to the transition for homopolymers; short-range repulsive interactions between different blocks have a negligible impact on the effective scaling exponents: they are the same regardless of the blocks being globule and coil or ideal and swollen coils. 

\keywords {polymers, scaling, universal properties, numerical simulations}

\end{abstract}

\section{Introduction}

In recent decades, block copolymers have come to the forefront of polymer science as they open up a wide range of potential applications for finely tuned materials~\cite{Bates2017,hadjichristidis2003}. Those traditionally include adhesives \cite{Sajjad20} and coatings \cite{Leonardi21}, but also extend to new areas such as targeted drug delivery \cite{Meng09, Xu24, Singh15}, tissue engineering \cite{Kutikov15, Malik21}, nanotechnology \cite{Schultz05}, and many others. This vast interest is largely due to the fact that block copolymers combine two or more types of chemically distinct monomers, each of which forms a long segment \cite{hadjichristidis2003}. As a result, block copolymers in melts have the capability to self-assemble into lamellae, micelles, and more complex structures \cite{Cummins20,Chen19,DAMI2017} due to microphase separation.

{ In most practical applications, polymers are found in dense solutions or melts; some studies suggest that understanding the size scales of individual molecules can help predict the pore size in the final product \cite{DAMI2017} since the characteristic size measured in a dilute solution is proportional to the size of the pore after self-assembley. Nevertheless, dilute solutions have also got a limited application. Examples of copolymers in a dilute solution are star-like micelles \cite{Hamley} and unimolecular micelles~\cite{GROMADZKI,Kelly22}. Another important point of dilute solutions is the capability to study the properties of a single molecule since  the interactions between molecules can be neglected \cite{Burchard}.}

One of the properties that characterizes the coil size and can be measured in experiments is the gyration radius. In the case of homopolymers, it is known that it increases with molecular mass according to the scaling law \cite{desCloiseaux,deGennes}:
\begin{equation}
    R_g^2 \sim N^{2\nu},
\end{equation}
where $N$ is a polymerization degree and $\nu$ is a Flory scaling exponent that does not depend on the microscopic details. In a poor solvent, this exponent is equal to, $1/d$ and in a good one to $3/(2+d)$. { In is important to note that Flory formula for the exponent in three-dimensional space gives an approximate value that is a bit bigger than the predictictions of  renormalization group, that is $0.5882(11)$ \cite{Guida98} or Monte Carlo simulations --- $0.587597(7)$\cite{Clisby10}. Experimental measurements of the scaling exponent for the radius of gyration, calculated for the five most common flexible linear polymers, range between 0.61 and 0.546~\cite{Fetters1994}. For DNA, using an atomic force microscope, the value of 0.585(14) was received \cite{Ercolini07}}.

However, in the case of copolymers, the story of scaling exponents is much more interesting~\cite{Holovatch97,Tanaka76} because usually the interaction between monomers and solvent molecules is preferable for one block rather than for the other one. It is a well-established fact in the literature that in the case of copolymers, each of the blocks is governed by its own exponent, while a third exponent governs the distance between the centers of masses of the blocks \cite{Tanaka76,olaj1998,Joanny84,Bendler77,Douglas87,vlahos1994,Mcmullen89,molina1995,Haydukivska19,Haydukivska24}. Unlike in the case of homopolymers, the experimental study of size characteristics for copolymers is also a hard task~\cite{Tanaka76,Tanaka77,Tanaka79}. Thus most of the research was done using theoretical approaches. 

The majority of the investigations of size characteristics of diblock copolymers were done either within the framework of the continuous chain model \cite{Douglas87,vlahos1994,Joanny84,Mcmullen89,Haydukivska19,Haydukivska24} or Monte Carlo simulations~\cite{molina1995,Douglas87,olaj1998,Haydukivska19,Haydukivska24,Zifferer2010} often making a comparison between those. Both analytical and numerical studies confirm the existence of the scaling exponents: one for each of the characteristic scale lengths. Two of those exponents are either $1/2$, which corresponds to the polymer in theta solvent, or {$\approx 3/(2+d)$}, known for the good solvent, and the third one is somewhere in between them. Moreover, it appears that the effective interaction between the monomers on different blocks does not affect the scaling exponents \cite{vlahos1994,olaj1998}. The conclusion also holds for more complex structures \cite{Haydukivska19,Haydukivska24,Zifferer2010,Vlahos1992,rubio2000}.

A notable point in all of those works is that they consider ideal blocks alongside swollen ones rather than collapsed ones, somewhat equating polymers in theta and poor solvents. In the case of a continuous chain model, this is true since a three-point interaction that describes a poor solvent has an upper critical dimension of three \cite{Kholodenko,Ohno}. As a result, those approaches account only for the influence of the repulsive interactions. 

In particular, the influence of attractive interactions on copolymers was studied~\cite{Oever,Singh22,Theodorakis}, showing the existence of the coil-to-globule transition for copolymers. However, to the author's knowledge, scaling properties for copolymers with attractive interactions were not studied. Thus, the aim of the present work is to study these properties along with universal shape characteristics of asphericity and prolatness~\cite{Aronowitz}:
\begin{eqnarray}
   \langle A_d\rangle=\frac{3}{2}\left\langle\frac{{\rm Tr}\, \hat{\bf S}^2 }{({\rm Tr}\, {\bf S})^2} \right\rangle,\label{Ad} \\
   \langle P_d\rangle=27\left\langle\frac{{\det}\, \hat{\bf S}^2 }{({\rm Tr}\, {\bf S})^2} \right\rangle,\label{Pd}
\end{eqnarray}
where $ {\bf S}$ is the gyration tensor, $\hat{ {\bf S}}= {\bf S}-\overline{\mu}{\bf I}$ with $\overline{\mu}$ being an average eigenvalue and ${\bf I}$ being a unity matrix. 

{ The simplest case of a block copolymer is a diblock copolymer, consisting of two chains connected together. Originally synthesized in the mid-1950s, they still draw the attention of the scientific community, both in traditional \cite{Sajjad20,Leonardi21,BRADFORD} and in new applications \cite{Chen19}. In the present work, it was chosen due to its simplicity that allows us to study the influence of interactions most clearly, because the aim of the study is to comprehend the influence of the interactions between monomers of different blocks on their universal properties.}

The layout of the paper is as follows: it starts by introducing the model and method in section~\ref{MD}, followed by the results and their discussion in section~\ref{RND}, and finally closing with concluding remarks in section~\ref{Con}.

\section{Model and methods} \label{MD}

\begin{figure}
    \centering
    \includegraphics[width=65mm]{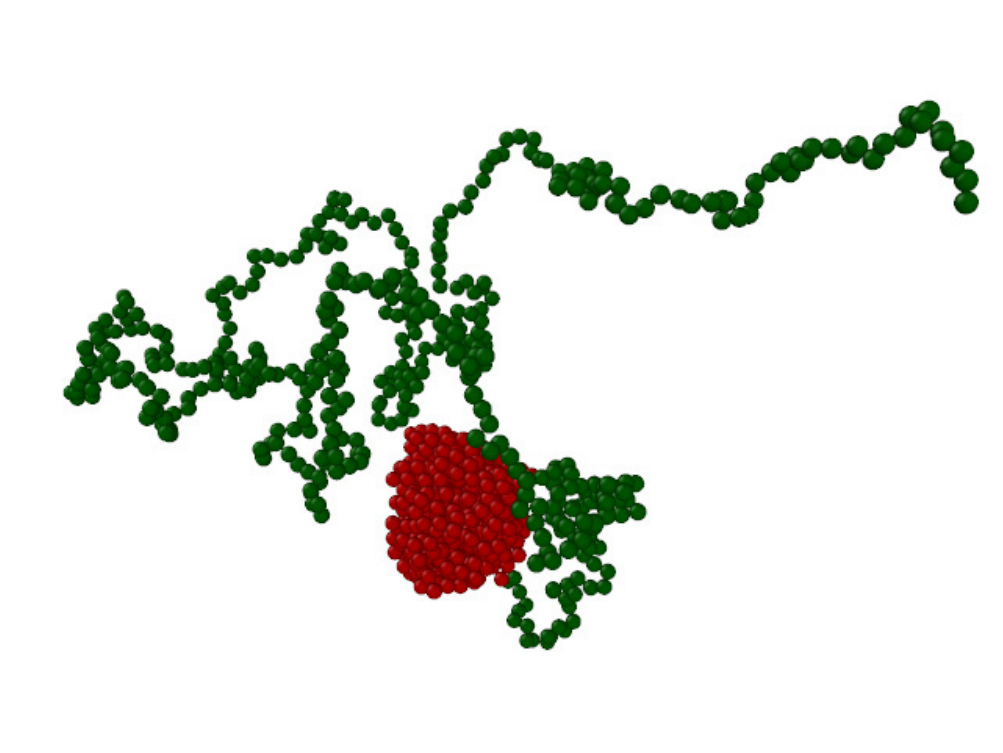}
    \includegraphics[width=65mm]{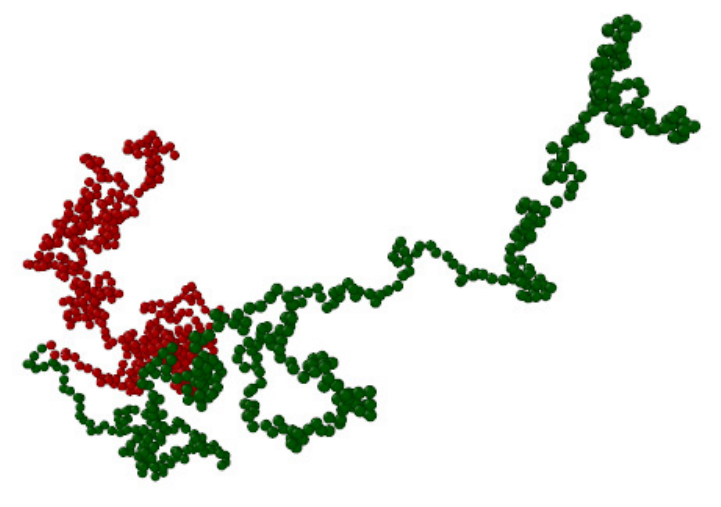}
    \caption{(Colour online) Simulation screenshot of diblock copolymer.}
    \label{scheme}
\end{figure}

The bead-spring coarse-grained model with two \cite{grest1987} types of beads (see figure~\ref{scheme}) is considered. The molecule contains equal parts of both. The beads are connected into the chains by springs described with the finitely extensible
nonlinear elastic (FENE) potential:
\begin{equation}
 V^{\mbox{\tiny FENE}}(r)=- 0.5kr_0^2\,\ln{[1-(r/{r_0})^2]}.
\label{fene}
\end{equation}
The interaction between nonconnected beads is introduced by the Lennard-Jones potential that was shifted and truncated, known as a Weeks-Chandler-Anderson (WCA) interaction:
\begin{eqnarray}
&& V^{\mbox{\tiny WCA}}(r) = 4\epsilon_{LJ}\left[
(\sigma_{LJ}/ r)^{12} - (\sigma_{LJ} /r)^6 - (\sigma_{LJ}/ r_{\rm cut})^{12} +(\sigma_{LJ} /r_{\rm cut})^6
\right]\theta(r_{\rm cut}-r),
\label{wca}
\end{eqnarray}
where $r$ is the distance between the centers of the beads with diameter $\sigma_{LJ}$, $\epsilon_{LJ}$ is an energy scale, and the constants are $k=30\epsilon_{LJ}/\sigma_{LJ}^2$ and $r_0=1.5\sigma_{LJ}$. $\theta$ is a Heaviside step function, and $r_{\rm cut}$ is a cutoff distance for the LJ potential. It is $2.5 \sigma_{LJ}$ for the type one beads, corresponding to the ``poor'' solvent, and has both attractive and repulsive parts of the Lennard-Jones interaction. The second type of beads corresponds to the ``good'' solvent, and thus only the repulsive part of the interaction, the cutoff distance for it, is $2^{1/6}\sigma_{LJ}$. Interaction between the types is set to be either only repulsion or both repulsion and attraction. { Symmetric diblock copolymers of length up to $900$ beads in a molecule (50/50 split between the types)  are considered.}
In total these potentials are known as Kremer-Grest potentials~\cite{grest1986}  $V^{\mbox{\tiny KG}}(r)=V^{\mbox{\tiny
FENE}}(r)+V^{\mbox{\tiny WCA}}(r)$.

The simulations were performed using the Large-scale Atomic/Molecular Massively Parallel Simulator (LAMMPS) \cite{lammps}, which solves a set of Newton's equations of motion implementing the velocity-Verlet algorithm with the iteration step $\Delta t = 0.005\tau$. In each run, the temperature $T$ was kept constant due to the introduction of the Langevin dumping term with the coefficient
$\zeta=0.5\,m\tau^{-1}$, where $\tau = \sqrt{m\sigma^2/\epsilon}$ is the LJ time unit and $m=1$ is monomer mass. The simulations were performed for a number of different fixed temperatures, both below and above the theta temperature.
The periodic boundary conditions in all three dimensions for the cubic box were implemented during the simulation. 

Each simulation box contained $27$ molecules, with the interactions between the molecules turned off to describe dilute solution conditions.{ Simulations run time was not less than three auto-correlation times plus an extra $10^7$ steps, with the latter part used for the calculation of the observable. Note that the observables were calculated after every $10^4$ steps, making in total at least $1000$ different samples per one chain. An average value is calculated  over an ensemble of all samples from all chains.}

\section{Results and discussion} \label{RND}

The gyration radius of the whole molecule is calculated using a definition:
\begin{equation}
    \langle R_g^2 \rangle = \big\langle\frac{1}{N}\sum_{i=1}^N (\vec{r_i}-\vec{r}_{\rm CM})^2 \big\rangle,
\end{equation}
where $N$ is a number of beads and $\vec{r_i}$ is a coordinate of a bead $i$ and $\vec{r}_{\rm CM} $ is the position of the center of mass. $\langle \ldots \rangle$ defines an ensemble average.

\begin{figure}[t!]
    \centering
    \includegraphics[width=100mm]{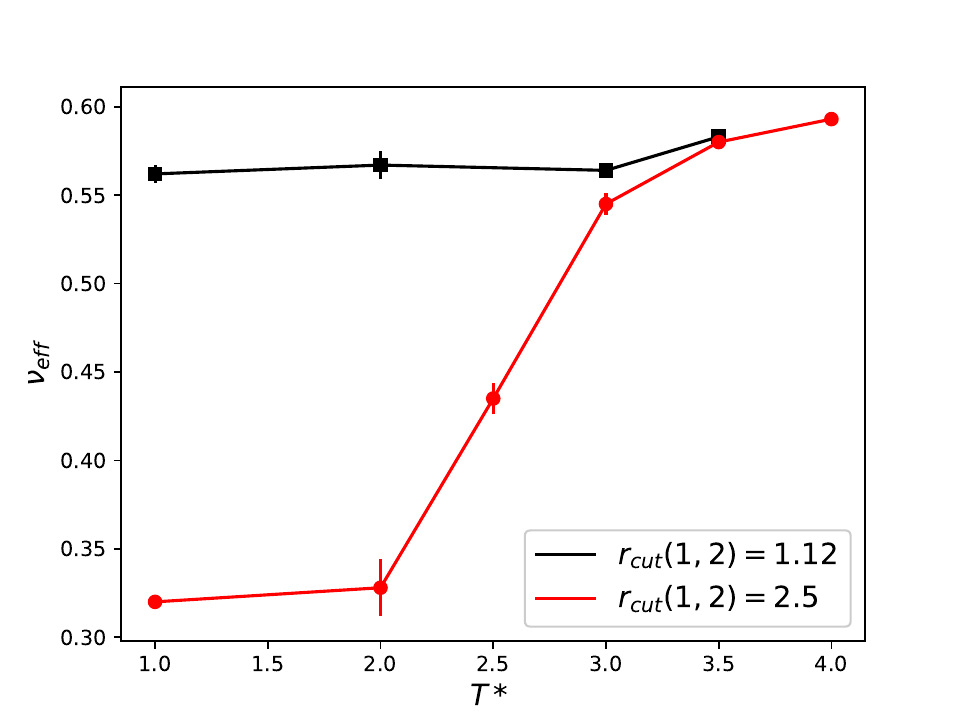}
    \includegraphics[width=100mm]{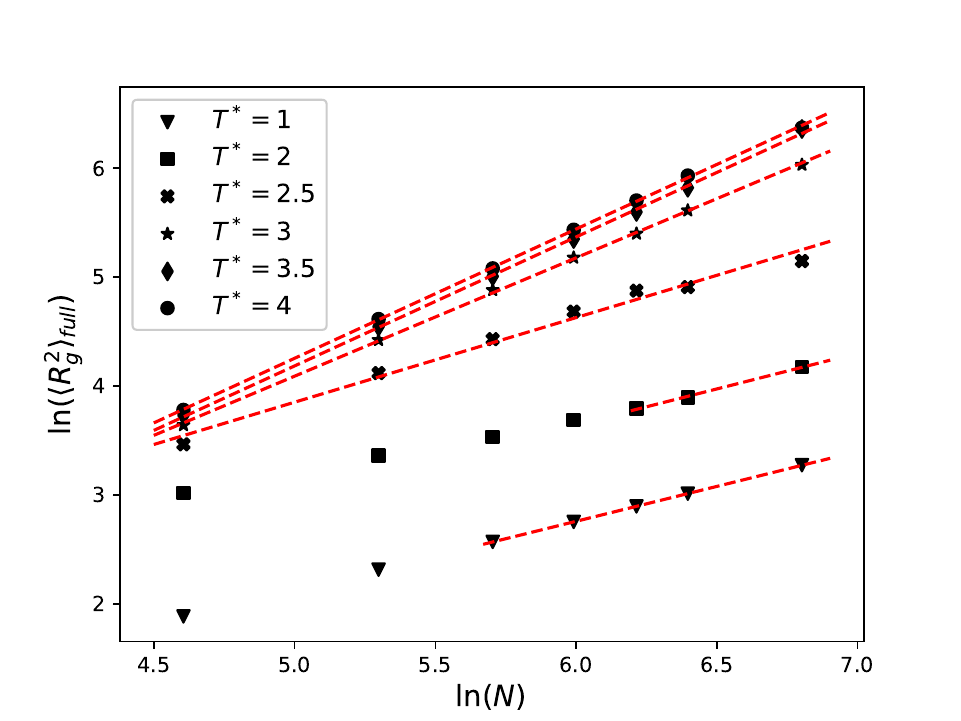}
    \caption{{(Colour online) Top: effective scaling exponent as a function of the dimensionless temperature~$T^*$. Bottom: gyration radius of diblock copolymer as a function of the total number of beads in a log-log scale for a case of $r_{\rm cut}=2.5$}.}
    \label{nueff}
\end{figure}

As was mentioned in the introduction, in general, for copolymers, there are three rather than one scaling exponent; however, since in experiments they are next to impossible to measure, it is more practical to consider an effective scaling exponent~\cite{Zifferer2010,Haydukivska24}, which, in a crude manner, can be calculated by simply looking at a slope of the data for the gyration radius versus the number of beads in a log-log scale. Such a calculation can be easily done using the data from either simulations or experiments. Results for such calculations for the data collected in simulations in the present work are presented in figure \ref{nueff}, {where the figure on top contains calculated values of the effective exponents, while at the bottom one, the corresponting fittings in the log-log scale are provided for a diblock copolymer with attractive interaction between the blocks ($r_{\rm cut}=2.5$). To test the accuracy of the fit, a quantity
	 $$1-\frac{\sum(R^2_{g,\,\rm{sim}}-R^2_{g,\,\rm{fit}})^2}{\sum(R^2_{g,\,\rm{sim}}-\overline{R^2_{g,\,\rm{sim}}})^2},$$
where $R^2_{g,\,\rm{sim}}$ is the data from simulation, $R^2_{g,\,\rm{fit}}$ is the data predicted by the fit and $\overline{(\ldots)}$ denotes an average over all samples, is considered for the fit. The smallest value of this quantity received was $0.985$ for a case of $r_{\rm cut}=2.5$ and temperature $T^*=2.5$}. As the temperature increases, the exponent increases for both considered cases. Here, in the case of an attractive interaction present between the monomers of different types ($r_{\rm cut}=2.5$), we see a classical transition from a ``poor'' to a ``good'' solvent. { Note,  however, that in the present work we are interested only in the fact that both ``poor'' and ``good'' solvent behaviors are present. The details of the transition require a deeper analysis and a second virial coefficient calculation, which is not the subject of this study.}

A more interesting behavior is observed for the case when only the repulsive interaction is considered between the monomers of different types ($r_{\rm cut}=1.12$). At the point of $T^*=1$, the scaling exponent for the part with an attractive interaction (see red beads in figure~\ref{scheme} on the left) is characterized by the scaling exponent $1/3$ typical of a polymer in a ``poor'' solvent, while at the point of  $T^*=3$ it is very close to $1/2$ (see red beads in figure~\ref{scheme} on the right) undergoing an expected transition. However, the effective exponent remains the same until both parts reach a good solvent behavior. The exponent for $T^*=1$ is $\nu_{\rm eff}=0.562(5)$ and for $T^*=3$ it is  $\nu_{\rm eff}=0.564(3)$. Both are in good agreement with the results from MC simulations that considered a case of random walk plus self-avoiding walks models of copolymer~\cite{Zifferer2010,Haydukivska24}.

\begin{figure}[t!]
    \centering
    \fontsize{14}{18}\selectfont
    \includegraphics[width=65mm]{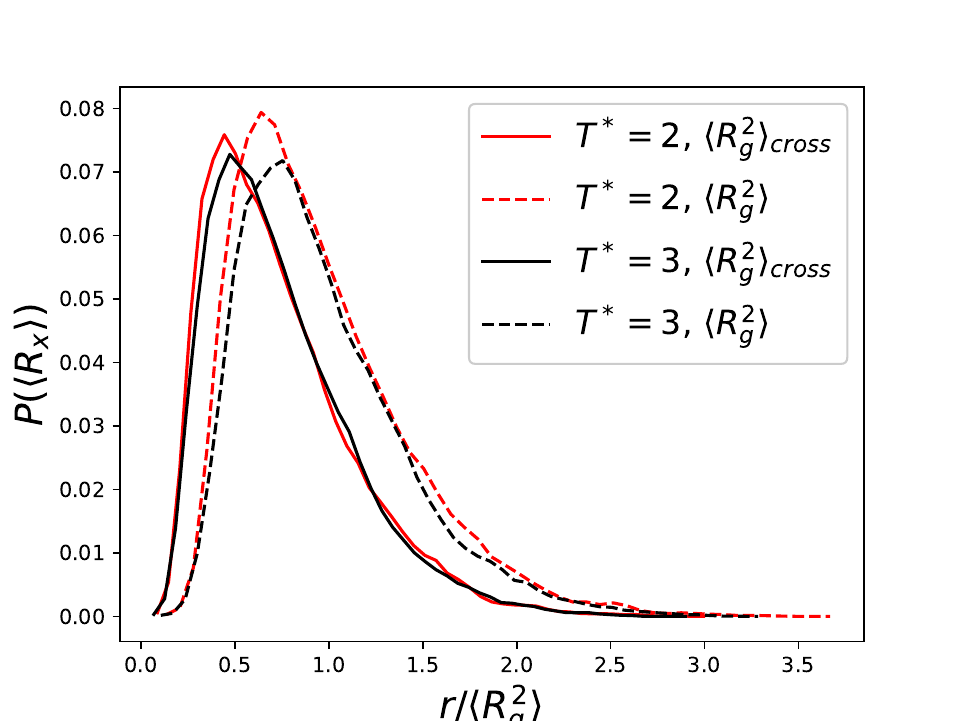}    
    \includegraphics[width=65mm]{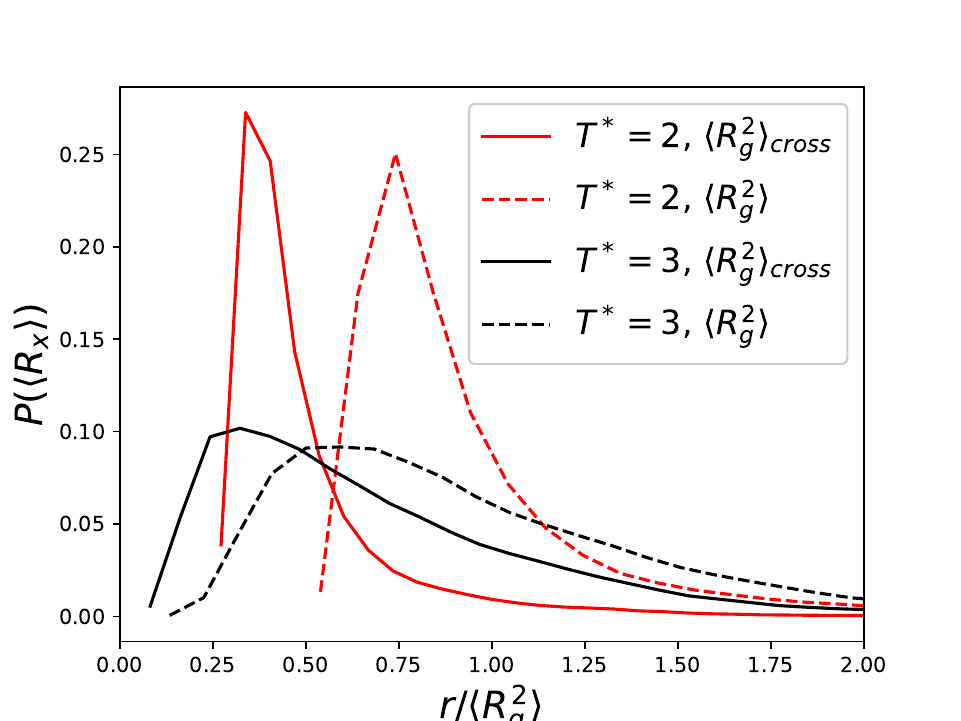}
    \includegraphics[width=65mm]{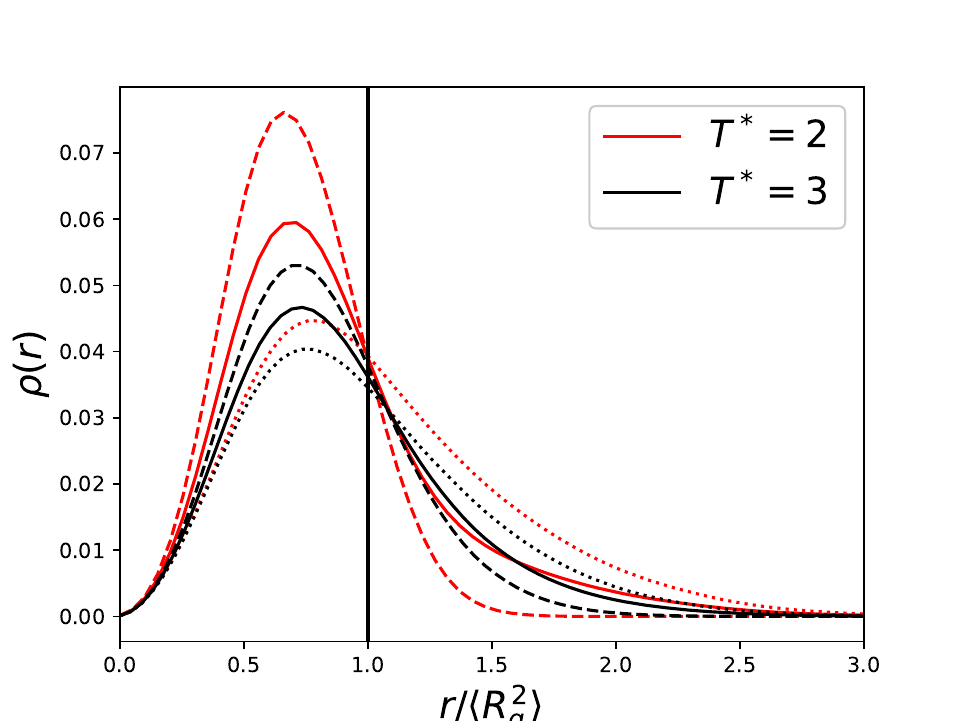}    
    \includegraphics[width=65mm]{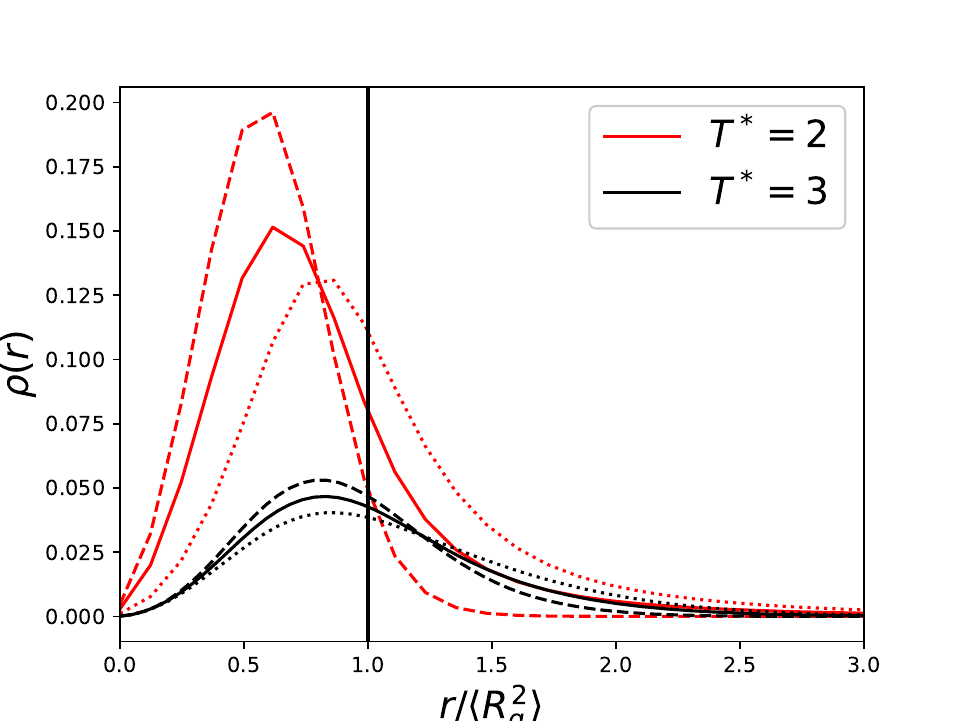}
    \caption{(Colour online) Top: distributions $P(R_x)$ as a function of observables normalized by the mean value of the gyration radius. The plots are given for $r_{\rm cut}=2^{1/6}$ (on the left) and for $r_{\rm cut}=2.5$ (on the right) for two different temperatures. Bottom: monomer density distribution function $\rho(r)$ as a function of distance from at the center of mass normalized by the mean value of the gyration radius. The plots are given for $r_{\rm cut}=2^{1/6}$ (on the left) and for $r_{\rm cut}=2.5$ (on the right) for two different temperatures.{ In the bottom plots solid lines show distributions for the whole molecule, while ``red'' beads and ``green'' beads are represented by dashed and dotted lines, correspondingly.}}
    \label{d25}
\end{figure}

So far, we looked separately at effective exponents and those that described the scaling of each block; however, it is important to note that in general, gyration radius can be presented as:
$$\langle R_g^2\rangle=\frac{N^2_{\rm{red}}}{N}\langle R_{g,\,\rm{red}}^2\rangle+\frac{N^2_{\rm{green}}}{N}\langle R_{g,\,\rm{green}}^2\rangle+\frac{2N_{\rm{red}}\,N_{\rm{green}}}{N}\langle R_{g,\,\rm{red}}\,R_{g,\,\rm{green}}\rangle.$$
 Each of those terms are characterized by its own scaling exponent. The first two are trivial and were mentioned above, and the third is often close in value to the effective one \cite{Zifferer2010,Haydukivska24}, thus not really providing any additional information. That is why in order to look a bit deeper into the conformational properties of the model, the largest system of 900 monomers is considered at two temperatures $T^*=2$ and  $T^*=3$: just below the transition and at the point of transition. 

To begin with, let us compare the distributions of $\langle R_{g,\,\rm{red}}\,R_{g,\,\rm{green}}\rangle$ and the corresponding gyration radii (see the top panels in figure \ref{d25}). Here the distribution of the gyration radius is given with a dotted line. Note that on the left-hand panel ($r_{\rm cut}=2^{1/6}$), distributions for different temperatures are quite similar in shape to each other and to the distributions of the mixed contributions, with the latter being only slightly shifted to the left. This indicates that for a case of only repulsive interaction between monomers of different types, the squared distances between them are the leading contributions to the gyration radius. From the samples of the trajectories (see figure~\ref{scheme}) it can be seen that an increase of the temperature also leads to some swelling in the red part of the molecule. The overall distances between red and green remain rather similar due to the green part being mostly unaffected by temperature, with only a slight change in size from $1.01\langle R^2_g\rangle_{\rm chain}$ to $1.05\langle R^2_g\rangle_{\rm chain}$, where  $\langle R^2_g\rangle_{\rm chain}$ is a gyration radius of a free chain of the same mass as the block. 

Note that on the right-hand panel $\langle R_{g,\,\rm{red}}\,R_{g,\,\rm{green}}\rangle$ provides a smaller contribution to the gyration radius for lower temperature and a significant contribution for a higher temperature, since all parts of the molecule undergo a transition due to the presence of interaction between the blocks. 

Another way to look inside the molecule is by calculating the monomer density distributions (see the bottom row in figure~\ref{d25}). On the left, results for a copolymer with $r_{\rm cut}=2^{1/6}$ for interactions between different types are provided, and on the right, results are presented for a copolymer with  $r_{\rm cut}=2.5$. The ``$x$''-axis is normalized by a gyration radius for a corresponding temperature. Note that for all cases the ``red'' monomers (dashed lines) are much closer to the center of mass, as expected; however, an overall change in monomer densities on the left-hand panel is smaller than on the right (solid lines). In fact, for a temperature $T^*=2$ on the left-hand panel, 70\% of monomers are closer to a center of mass than the average (gyration radius),  whereas on the right, it is 80\%. An increase of the temperature leads to a decrease of those numbers to the corresponding 68\% and 69\%. This expectedly indicates a much stronger change when there is an attractive interaction between monomers of different types, but there is only a small change when there is only a repulsion interaction. 

It is also noteworthy that, in the case of $r_{\rm cut}=2^{1/6}$ (left-hand panel), the most likely distance of the ``red'' and ``green'' monomers from the center of mass (position of a maximum on the distribution) is nearly the same, while, for lower temperatures, there is a noticeable difference on the right. 

\begin{figure}[t!]
    \centering
    \includegraphics[width=65mm]{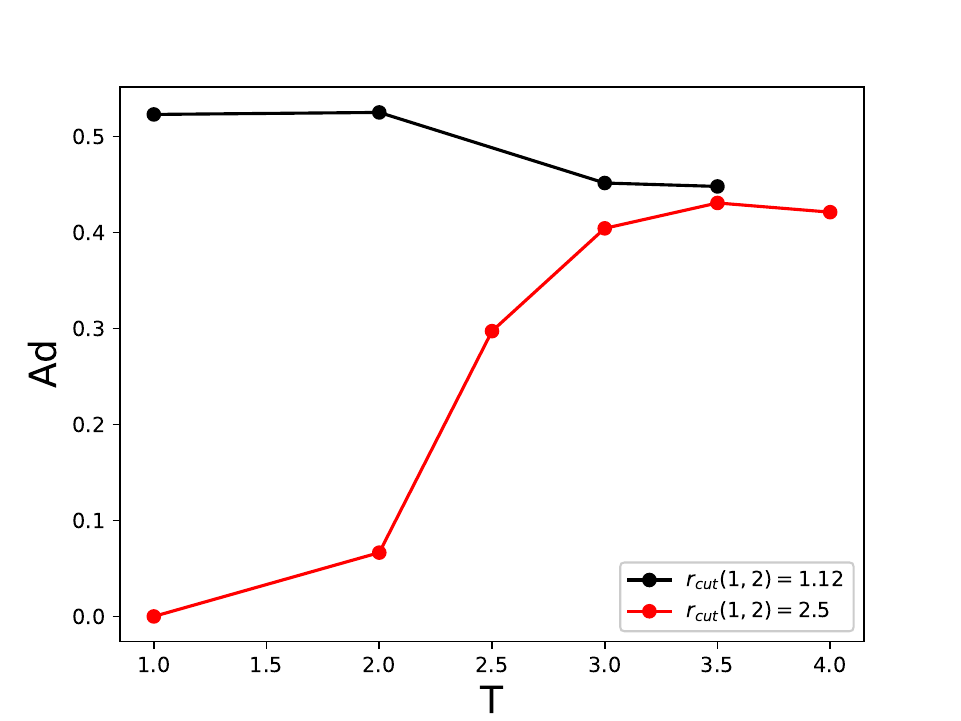}    
    \includegraphics[width=65mm]{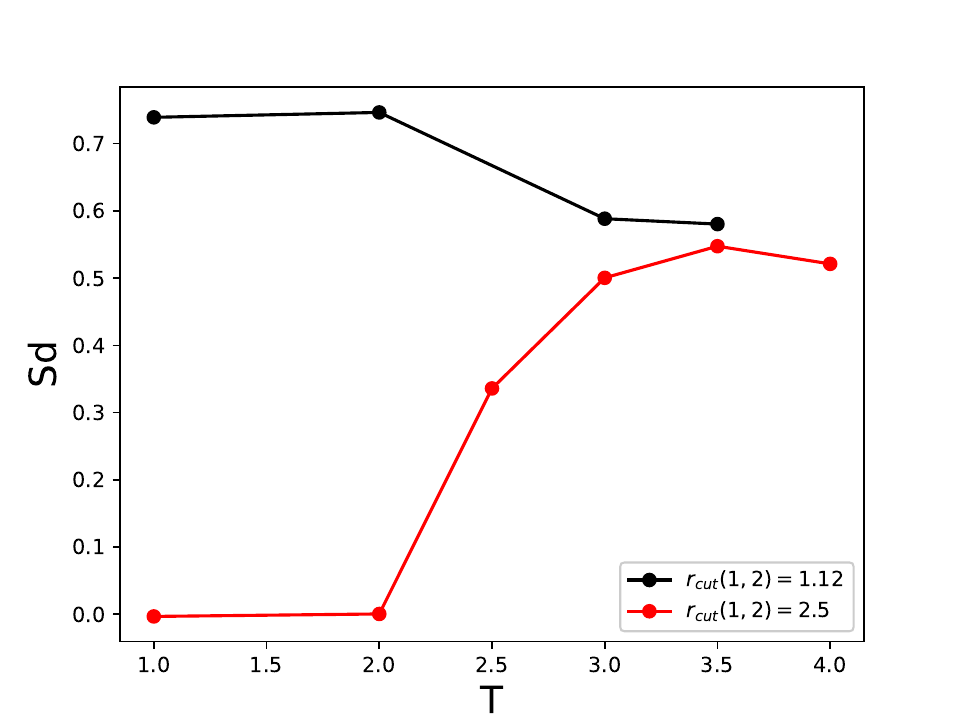}
    \caption{(Colour online) Left: asphericity as a function of the dimensionless temperature $T^*$. Right:~prolatness as a function of the dimensionless temperature $T^*$.}
    \label{shape}
\end{figure}

Having explored the spatial distribution of the monomers, we now turn our attention to the overall shape of the system, quantified by its asphericity and prolatness. The first one distinguishes between spherical and rod-like shapes, and the second one distinguishes between oblate and prolate ellipsoids. Here, the results were calculated using the finite size approximation $A_d(N)=A_d+BN^{-1/2}$ and are provided in figure \ref{shape}. Note that for high temperatures, both characteristics reach typical values for a polymer in a ``good'' solution, while predictably for a copolymer with attractive interaction between the blocks, there is a transition from spherical to elliptical shape. Once more, a copolymer with repulsive interaction between blocks exhibits a more intriguing behavior, since the shape characteristics exhibit somewhat higher values at lower temperatures.
 This indicates a more prolate shape, caused by the repulsive interaction that stretches the solvable block. Also note that for $T^*=2$ and  $T^*=3$, the  scaling exponents are the same, but the shapes are different, pointing to a difference in the impact of a globule and ideal chains on the shape characteristics. 

To close this consideration, let us point out that considering one block as ideal instead of collapsed in case when interaction between blocks is repulsive, it does not affect the scaling behavior. This allows continuous chain models and Monte Carlo models to be powerful methods in studying those copolymers in a dilute solution. However, block-to-block attraction cannot be treated in the same way.

\section{Conclusions}\label{Con}

In the present work, we consider a bead-spring model of diblock copolymer in a dilute solution, in order to understand the influence of different types of monomer-monomer interactions onto its size characteristics, in particular, the effective scaling exponent. One part of the diblock is considered to be in poor solvent while the other one is in good.  

The main attention is paid to the differences between monomers of different types, and whether there is an attractive interaction between them. The influence of these interactions is studied as a function of the temperature in the range that covers a theta point. 

{As expected, we find a strong influence of an attractive interaction on the effective scaling exponent, with behavior being similar to homopolymers. On the other hand, in the case when there is only a repulsive interaction between the monomers of different types, the effective scaling exponent changes only slightly around the theta temperature and stays the same as the temperature drops down. This is caused by an expanded crown of monomers for which the solvent is good that is much more extended than the other type. Numerical values received in simulations are in good agreement with Monte Carlo simulations conducted for the macromolecules modelled as random walk (theta solvent) plus self-avoiding walk (good solvent). 

On the other hand, the shape characteristics reach different values when one of the blocks is in either ``poor'' or theta solvent, which indicates that only for the case of effective scaling exponents, the random-walk-like and colapsed states can be equated. This, however, shows that analytical calculations of effective scaling exponents within a continuous chain model can be of potential scientific interest. 

In addition, at this stage, a connection between dilute solution properties and the properties in a concentrated state are implied in the literature, but to the author's knowledge the evidence for the connection or the lack thereof is sparce, and thus it is the author's hope that this small work may be a building block for further studies.} 

\section*{Acknowledgements}
	The author would like to acknowledge support from the National Science Center, Poland (Grant~No.~2018/30E/ST3/00428) and the computational time at PL-Grid, Poland.

\bibliographystyle{cmpj}
\bibliography{name}

\ukrainianpart

	\title{Властивості скейлінгу для диблок кополімерів: молекулярна динаміка}
\author{Х. Гайдуківська\refaddr{label1,label2}}
\addresses{
	\addr{label1} Iнститут фiзики конденсованих систем Нацiональної академiї наук України,
	79011, м. Львiв, вул. Свєнцiцького, 1, Україна
	\addr{label2} Iнститут Фiзики, Сiлезький унiверситет, вул. Першого полку пiхоти 75, 41-500 Хоржув, Польща
}

\makeukrtitle

\begin{abstract}
	\tolerance=3000%
	У роботі вивчається вплив взаємодії між мономерами на скейлінгові показники та характеристики форми полімеру в розчині з використанням методу молекулярної динаміки. Зокрема, вивчається вплив взаємодії між мономерами на різних блоках на конформаційні властивості полімеру при різних температурах. Результати показують, що у випадку макромолекул з притягальною взаємодією між блоками зі зростанням температури спостерігається перехід подібний до характерного для однорідного полімеру; натомість, коли між блоками є лише відштовхувальна взаємодія, її вплив є нехтовно малим, а ефективний скейлінговий показник буде таким же, незважаючи на те, чи блоки є глобула і клубок, чи це ідеальний і набухлий клубки.
	\keywords полімери, скейлінг, універсальні властивості, числове моделювання
\end{abstract}

\end{document}